\documentclass{article}

\usepackage{arxiv}

\usepackage[utf8]{inputenc} 
\usepackage[T1]{fontenc}    
\usepackage{hyperref}       
\usepackage{url}            
\usepackage{booktabs}       
\usepackage{amsfonts}       
\usepackage{nicefrac}       
\usepackage{microtype}      
\usepackage{lipsum}		
\usepackage{graphicx}
\usepackage[numbers]{natbib}
\usepackage{doi}

\title{Patient-Specific 3D Printed Dynamic Preoperative Planning Models in Modern Medicine}

\author{Keshav Jha \\
  Duke University\\
  3361 S Beverly Pl \\
  Chandler, AZ 85248 \\
  \texttt{keshav.jha@duke.edu} \\
  \And
  \href{https://orcid.org/0000-0000-0000-0000}{Joseph Mayer} \\
  Mayo Clinic Arizona\\
  5711 E. Mayo Blvd. Support Services Building\\
  Phoenix, AZ 85054 \\
  \texttt{mayer.joseph@mayo.edu} \\
}

\renewcommand{\shorttitle}{Dynamic 3D Printing}

\hypersetup{
pdftitle={Dynamic 3D Printing},
pdfsubject={physics.med-ph},
pdfauthor={Keshav Jha, Joseph Mayor},
pdfkeywords={Preoperative planning, 3D-printed, dynamic models, multi-part, removable, anatomical models, patient-specific, segmentation},
}

\begin{document}
\maketitle
\begin{abstract}
	Three-dimensional (3D) printed preoperative planning models serve a critical role in the success of many medical procedures. However, many of these models do not portray the patient's complete anatomy due to their monolithic and static nature. The use of dynamic 3D-printed models can better equip physicians by providing a more anatomically accurate model due to its movement capabilities and the ability to remove and replace printed anatomies based on planning stages. A dynamic 3D-printed preoperative planning model has the capability to move in similar ways to the anatomy that is being represented by the model, or reveal additional issues that may arise during the use of a movement mechanism. The 3D printed models are constructed in a similar manner to their static counterparts; however, in the digital post-processing phase, additional care is needed to ensure the dynamic functionality of the model. Here, we discuss the process of creating a dynamic 3D-printed model and its benefits and uses in modern medicine. 
\end{abstract}


\section{Introduction}
Since its initial creation in 1983, 3D printing’s uses have expanded dramatically along with the technology \cite{hall2015birth}. The uses of 3D printing have begun to expand into the medical sphere, and its use has expanded into the realm of prosthetics, implants, presurgical planning models, and more \cite{hurst2016healthcare}. Specifically, the use of 3D printing for presurgical planning models has revolutionized preoperative planning and allowed physicians to have greater confidence in the chosen procedure and how they plan on proceeding \cite{wake2017renal}. Although current 3D-printed models represent the basic static anatomy, they do not represent the complexity of how the body moves, and multiple models may be needed to highlight different aspects of the physical anatomy.

To overcome this, dynamic, multi-part 3D-printed models can be used. Dynamic models have the ability to move, twist, and turn in paths resembling the specific anatomies, as well as the ability to come apart to reveal additional structures and how each structure interacts with one another. These functionalities can have a wide range of applications in preoperative planning and make preoperative printing more effective and efficient.

Here, we give an in-depth review of the creation of dynamic 3D-printed preoperative planning models, evaluate the advantages and disadvantages of different 3D printing technologies for producing dynamic models, and discuss the uses of dynamic models and how these models may be improved in the future.

\section{\textbf{Process of Creating a Dynamic Model}}
\label{sec:headings}

The general process of creating a dynamic model is identical to that of a static 3D printed model, with additional care being taken during the digital post-processing step to ensure proper model functionality. First, using medical imaging techniques, an image set is gathered of the anatomy of interest. The images are segmented, and a Standard Tessellation Language (STL) model is created, which can then be 3D printed. Utilizing digital post-processing software, the model is refined, and care is taken to ensure that there is no interference between the parts that will be dynamically interacting. Coloring, labels, and identifiers applied here play a fundamental role in allowing physicians to easily identify various unique structures within the model. The model is then printed and post-processed through techniques such as wet sanding and clear coating to further refine the model and make the moving parts easier to interact with each other.  

\subsection{Image Acquisition
}
The process of creating an anatomically accurate 3D model requires the use of any type of volumetric imaging that can differentiate between tissues, such as computed tomography (CT) and magnetic resonance imaging (MRI) \cite{wake2017renal}. CT images are the most commonly used due to the ease of segmenting many types of critical anatomies. CT images also allow for superior tissue differentiation due to signal-to-noise ratio (SNR) and spatial resolution. When creating a dynamic model, image sections of a CT scan should be generated with isotropic voxels, essentially volumetric pixels. Generally, voxel sizes vary from 0.25 mm to 1.25 mm, and the optimal voxel size is dependent on the anatomy being represented. Larger voxels compromise print accuracy and quality for faster imaging and reduced exposure times, while smaller voxels require extensive digital post-processing and segmentation to ensure a usable model. The x- and y-dimensions of the voxel are determined by the detector size within CT machines, so they cannot be easily changed. On the other hand, the z-dimension, also known as “slice thickness,” can be easily adjustable, with many thin slices resulting in better-quality image sets \cite{chepelev2018rsna}.

Since dynamic models are meant to move freely, the interference between the moving parts should be minimized. Therefore, voxel sizes should generally be smaller to prevent rough edges on models and lower post-processing times on the final model. However, larger voxels can still be used to create a dynamic model in a quicker and less accurate manner, but additional digital post-processing may be necessary, and the dynamic nature of the model may suffer.

\subsection{Segmentation}
Acquired images that represent 3-dimensional anatomy as 2-dimensional images are stored as Digital Imaging and Communications in Medicine (DICOM) files. In order to be 3D printed, these files must be manipulated using a segmentation software such as MeVisLab (Mevis Medical Solutions, Bremen, Germany) or Mimics (Materialise, Leuven, Belgium) which allows for the creation of anatomical masks that can be turned into 3D digital models \cite{wake2017renal}. 

\begin{figure}[h]
    \centering
    \includegraphics[width=0.5\linewidth]{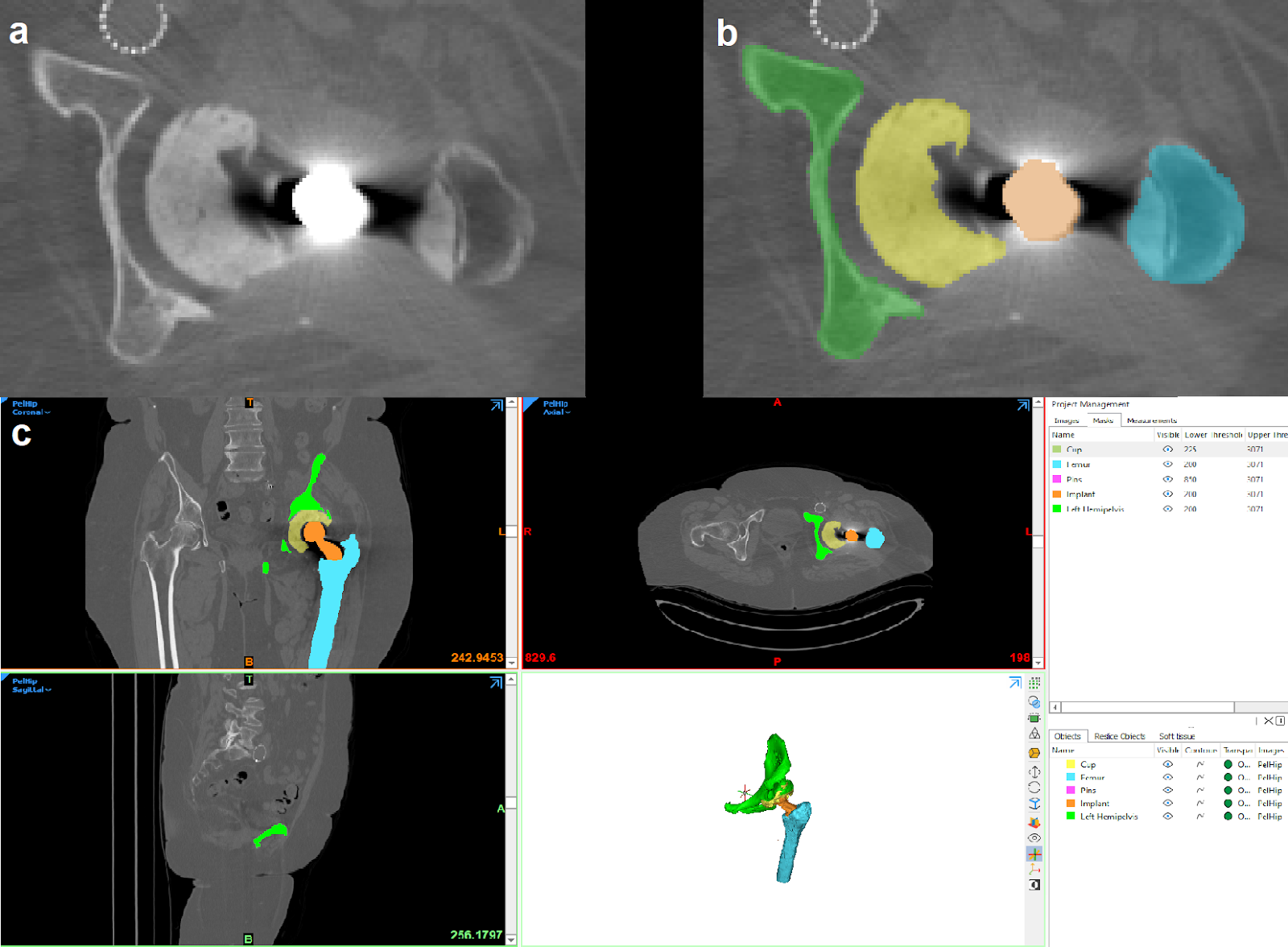}
    \caption{ Segmentation of an acetabular cup, femoral prosthesis, hemi-pelvis, and femur. Each anatomical part is segmented into a unique mask of varying colors to differentiate between each part. \textbf{(a)} Initial CT scan of the hemi-pelvis, acetabular cup, femoral prosthesis, and femur. \textbf{(b)} CT showing segmentation masks of the hemi-pelvis in green, the acetabular cup in yellow, the femoral prosthesis in peach, and the femur in cyan. \textbf{(c)} Materialise Mimics workspace showing the segmented anatomies of interest along with the initial digital model ready for post-processing.
}
    \label{fig:placeholder}
\end{figure}

Using methods of automated and manual segmentation such as region growing, thresholding, and manual mask editing, regions of interest (ROI) can be isolated for use in the creation of 3D digital and eventually physical models \cite{chepelev2018rsna, giannopoulos2016cvd}. Thresholding is an efficient tool to quickly isolate different anatomical structures based on their density, represented in Hounsfield units; however, holes may form in the mask due to areas of low signal intensity. After an initial global thresholding, some manual editing, such as adding or removing mask area, may still be required to acquire an accurate ROI and fill in any gaps that may be present. For more numerous gaps or thin masks, an automatic tool such as Smart Fill can quickly and efficiently fill in these gaps and create the desired mask. Region growing is an automated segmentation tool used to determine voxels of different anatomies by identifying voxel density, and can significantly reduce post-processing times and manual editing of the model by growing seed points based on the desired anatomies' density and connecting voxels \cite{mitsouras2015radiologist}. Once all of the anatomies of interest have been segmented into their respective masks, a digital model is created for further post-processing (Figure 1).

\subsection{Digital Post-Processing}
After the segmentation of the DICOM files is complete, the masks are converted into STL models and exported for digital post-processing. Most segmentation software packages automatically generate an STL file of the segmented anatomy using algorithms such as interpolation or pattern recognition that convert surfaces surrounding segmented anatomical structures into triangular facets. The number of triangular facets that make up the STL model is dependent on a multitude of factors, with the main one being the slice thickness of the CT image set, where thicker slices will produce larger voxels, which will lead to fewer large triangular facets in the model. STL models with too few triangular facets will lose their accuracy, while too many can lead to excessive computing and post-processing times \cite{mitsouras2015radiologist}.

These generated STL files, although directly compatible with 3D printers, generally need additional post-processing to achieve the desired results and ensure proper printability and functionality \cite{segaran2021preop, hangge2018aortic}. Commonly, much of the post-processing is performed in computer-aided design (CAD) software, such as Fusion 360 (Autodesk, San Rafael, California), Magics (Materialise, Leuven, Belgium), or 3-Matics (Materialise, Leuven, Belgium). Unwanted artifacts that may have been generated during the segmentation process, such as holes between triangular facets or inverted normals, can be fixed using these software packages. These software packages have automatic error repair tools that will deal with common modeling errors, but in some cases, manual repair tools may be needed to address various errors directly. To address rougher topography on the model, smoothing tools are effective at limiting harder edges and creating a more accurate and functional model. However, excessive smoothing can result in a loss of anatomical structure and accuracy \cite{mitsouras2015radiologist, segaran2021preop}.

For the creation of Dynamic models, these post-processing steps are the most critical to ensure that each unique anatomical mesh does not interfere with one another. In a location where two or more parts, which need to move independently of one another, meet, care must be taken to ensure that each unique mesh does not interfere with another mesh as this will cause the parts to be printed as a monolithic component, eliminating their ability to move independently of each other. Utilizing the smoothing, push-pull, or translation tools will allow for each mesh to be worked on to ensure that there is no overlap between two or more parts (Figure 2).

\begin{figure}[h]
    \centering
    \includegraphics[width=0.5\linewidth]{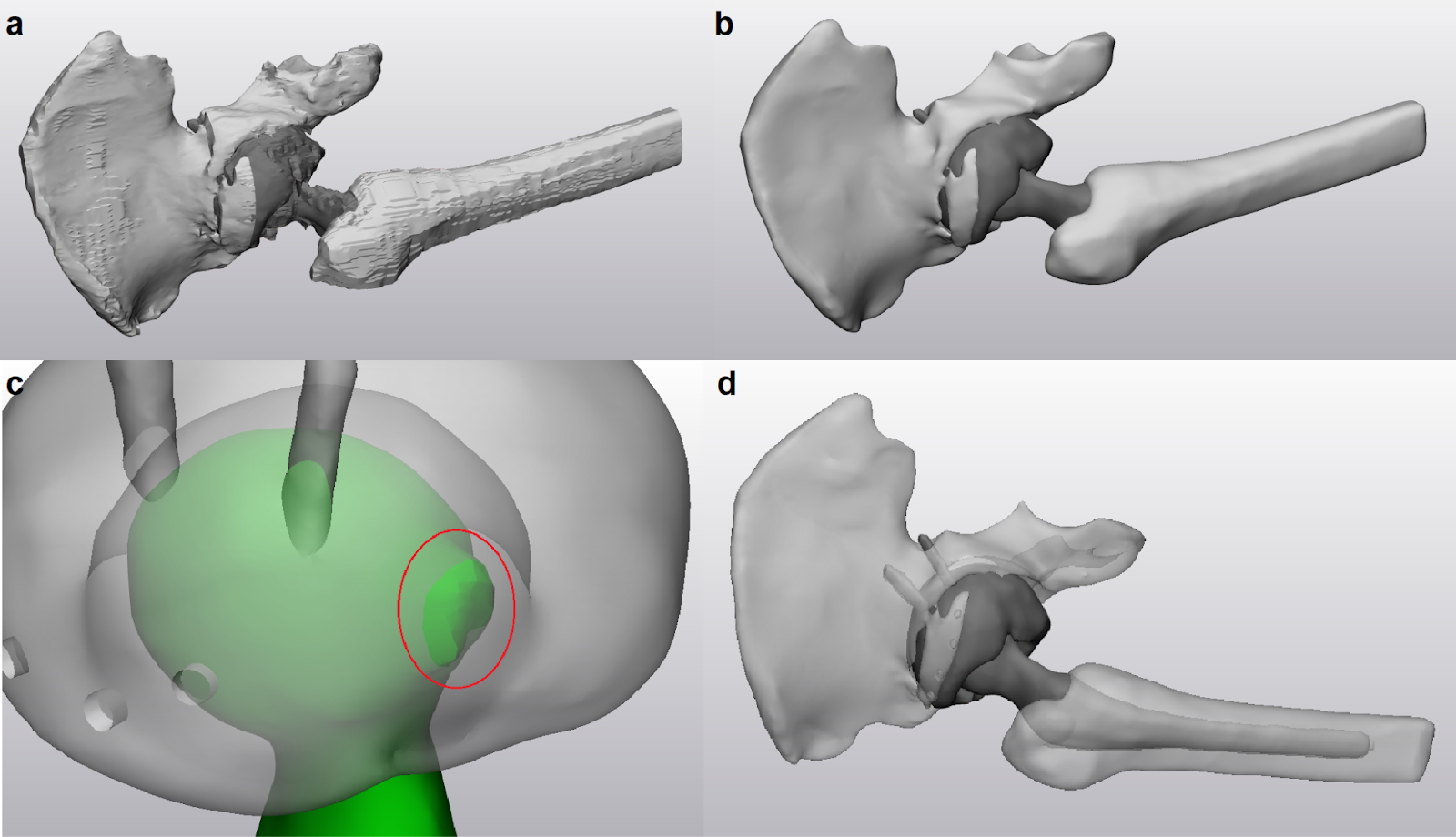}
    \caption{Post-processing of a digital model to ensure that the anatomies are correct and that there is no interference between multiple parts. \textbf{(a) }Initial rough digital model generated from the segmented masks. \textbf{(b) }Post-processed and smoothed models ready for slicing and printing. \textbf{(c) }Interference between the femoral prosthesis, seen in green, and the acetabular cup. This error will cause the parts to be printed as one and keep them from moving. \textbf{(d)} Fully completed model shown in transparent to verify that there is no interference between two parts that need to be dynamic.}
    \label{fig:placeholder}
\end{figure}

\subsection{3D Printing}
The completed digital model is imported into the respective slicing software for the 3D printer that will be used. When creating a 3D printed model, it is recommended that the resolution of the 3D printer be greater than that of the medical images used to create the model, ensuring that enough material will be deposited per layer to show the entirety of the anatomical structure. Furthermore, the layer thickness of a medical model should be around one-eighth of a millimeter, but can differ depending on the anatomy being modeled \cite{shilo2018future}.

There are many different methods of 3D printing, and consideration of multiple factors goes into choosing the best method for the model being produced. It is important to consider factors such as time, cost, availability of resources, and the use case of the model when choosing a printing method \cite{segaran2021preop, gokuldoss2017amprocess}. Here we will discuss the advantages and disadvantages of using the mainstream printing technologies when creating a dynamic model. Other 3D printing technologies, such as sheet lamination and directed energy deposition, are less commonly used in medical 3D printing and will therefore not be discussed here \cite{mitsouras2015radiologist, segaran2021preop}.

\subsubsection{Stereolithography}
Stereolithography (SLA), also known as VAT photopolymerization or digital light processing (DLP), is a widely used method of 3D printing. During the process of stereolithography, photosensitive liquid resin is cured in layers upon a printing bed by high-intensity UV light. Upon completion of the model, it is rinsed in a cleaning solution, usually Isopropyl alcohol, to remove any excess resin and left to cure inside a UV light chamber. Due to the use of a single material that is printed during the SLA process, only a single colored model can be produced. However, selective darkening can be applied to areas of interest by overexposing those areas to the UV curing light, which enhances the ability to view internal anatomies, such as nerves, tumors, and veins \cite{segaran2021preop}. 

Dynamic models are possible to make with SLA printers and are a relatively accurate and low-cost method to create them. However, SLA printers utilize a lattice structure, made of the same material as the rest of the part, to support the printed parts, which will need to be manually removed and may not be dense enough to support parts that are not monolithic, such is the case with dynamic models. SLA printing has the capability to create dynamic models and be relatively accurate and inexpensive; however, there are more time-efficient, accurate, and easier methods that could be used.

\subsubsection{Binder Jetting}
Binder Jetting (BJG) is a type of 3D printing technology in which a printing bed of powdered material, such as gypsum, polymers, metal alloys, or ceramics, is sprayed with a liquid binding agent. After each layer is sprayed, new powdered material is added and sprayed to create consecutive layers, resulting in a 3D model. BJG is a relatively inexpensive method of 3D printing, as many of the materials, including colored materials, are low-cost relative to other 3D printing technologies. The ability to print in various colors during the BJG process allows for color-coded anatomy models, along with differentiation of multiple anatomies to be created. Furthermore, support structures for models are not needed due to the unbound powder on the build plate, which acts as support for the model. Postprocessing of models entails cleaning off excess powder and “infiltration” of the model, which can yield deformable models \cite{mitsouras2015radiologist}. A disadvantage of the BJG process is that transparent models can not be created through this process, which eliminates the ability to create models to view internal anatomies. Models printed with this technology are also comparatively less durable to wear than those of other methods of printing \cite{segaran2021preop, revilla2018dentistry}.

BJG printing is a great choice for the production of dynamic models since the unbound powder in the print volume is easily removed and can support parts that are not monolithic. One thing to keep in mind, however, is that these models can be more fragile than other types of printing technologies, especially directly after the printing process is completed, and may not hold up in the long run when parts rub against each other.

\subsubsection{Material Jetting}
Material Jetting (MJ), also referred to as polyjet or multijet printing, jets a liquid photopolymer resin onto a build plate and cures the liquid with UV light to harden it into a solid model \cite{gonzalez2018extrusion}. The liquid resin is sprayed by multiple print heads, each assigned unique materials such as support material, clear resin, various colored resins, or elastomers. After each spray of the liquid and subsequent curing through UV light, the build plate is lowered layer by layer to form the model \cite{mitsouras2015radiologist}. One major advantage of MJ printing is that the support material for the model can be created with a gel or wax-like material that is easily removed in a solution or manually after the model has completed printing. The materials used for MJ are versatile and have a multitude of colors and properties available, which makes it suitable for the creation of detailed medical models. Furthermore, the materials can be mixed to enhance physical properties and create different textures such as skin or bone, which makes it useful for the creation of dynamic medical models due to the enhanced realism of the models created. However, MJ materials are relatively expensive and have expiry dates, which can cause failed prints if the material is out of date and has degraded \cite{mitsouras2015radiologist, segaran2021preop}. 

Due to the flexibility of MJ, it allows for the creation of durable dynamic models. Different textures, such as skin and bones can be printed, and the wide range of colors and material properties allows for a more detailed and differentiated model to be created. Furthermore, the easily removable support structures allow for the creation of movable interlocking parts such as skeletal joint systems. The MJ printing technology is perhaps one of the better choices for the production of dynamic models.

\subsubsection{Material Extrusion}
Material Extrusion, also known as Fused Filament Fabrication (FFF) or Filament Deposition Modeling (FDM), is a form of 3D printing in which heated material filament is extruded through a nozzle and onto a build plate layer by layer. The extruded material then cools and hardens to create a functional model. FFF is the most common form of 3D printing due to its ease of use, low cost, and introduction into the hobbyist community \cite{segaran2021preop, gonzalez2018extrusion}. Although this process is not widely used in the medical field due to its low quality and low resolution of models, it can be used to quickly create a model when the quality and accuracy are not a top priority \cite{segaran2021preop, revilla2018dentistry}.  Another benefit of the FFF printing method is the wide range of materials that are available for the printers. This benefit can allow for models to be printed in super strong, high-heat-resistant, or flexible materials.

The creation of dynamic models and movable parts is possible with FFF; however, extensive post-processing such as sanding of edges and applying protective coatings to increase durability would be needed. Most low-cost FFF printers utilize a lattice-like support structure that is printed in the same material as the model, so manual removal of this support structure would be needed to utilize this method for the production of dynamic models. However, some higher-end enterprise FFF printers utilize multiple nozzles that can print a support structure in a material that will dissolve in a caustic bath, allowing for the easy removal of the support material from tight spaces.

\subsubsection{Powder-Bed Fusion}
Powder-bed fusion is a form of 3D printing that includes selective laser sintering (SLS), selective laser melting (SLM), direct metal laser sintering (DMLS), electron beam melting (EBM), and Multi-Jet Fusion (MJF). These methods use high-powered lasers, electron beams, or heat-activated agents in order to melt and fuse material particles into a solid part. After a layer has been fused, a thin layer of powder is deposited and fused with the previous layer to create a functional 3D part. In some cases, models created by powder-bed fusion printers may not need support structures since the powder around the model supports the part that is being printed. Models created by this process are durable, and due to the unfused powder around the part doubling as support, models can be constructed using complex 3D geometries such as lattices to reduce weight and show internal structures of interest \cite{mitsouras2015radiologist, segaran2021preop}. Due to the various materials that are available for this type of printing, specifically metal alloys, this printing technology allows models to be used directly as medical implants. Post-processing of models varies depending on the material used for printing; for example, metals may need computer numerical control (CNC) milling or wire electrical discharge machining (EDM) to remove support structures and to achieve the desired finish.

Powder-bed fusion technologies are a great choice for the production of dynamic models, due to the wide range of materials available and the properties of those materials. Powder-bed fusion processes can also print in multiple colors and apply various textures to the parts to create a more useful model. Due to the nature of the unfused powder supporting the part, dynamic models can easily be created with minimal post-processing steps needing to be applied. 

\subsection{Post-processing of the 3D Model}
Many 3D printing technologies can produce an accurate and functional part after printing; however, some post-processing steps are often necessary to further refine the model and produce a fully functional and end-user desired part. All printing methods utilize support structures to keep the model steady on the build plate and to ensure that all parts, regardless of orientation, are printed fully and accurately. This extra material needs to be removed once the print is completed to show the full finished model. Some printing technologies, such as SLA and FFF, utilize a lattice structure that is made of the same material as the model to support the various parts. This lattice support needs to be manually removed by an operator with the aid of tools such as pliers and flush cutters, which can prove difficult in tight spaces such as joints and other small anatomical interfaces. Other printing technologies, such as powder-bed fusion and binder jetting, utilize the unfused powdered material as a support structure. This material can be reclaimed by the printer either automatically or through the use of a manual vacuum system. The Final type of support structure, mainly used in MJ but sometimes in higher-end FFF printers, is the use of dedicated support material that can be dissolved in a water or caustic bath. Powder-bed fusion and MJ printing support methods are the preferred methods for dynamic models as these supporting methods will allow for tight, dynamic joints to be printed, while the support material can be easily and quickly removed.

Once the support structures have been removed from the printed model, additional post-processing steps can be performed depending on the desired outcome and use of the model. Wet sanding and clear coating can be performed on the model to ensure a smooth and durable surface or to make clear resins fully transparent. In the case of dynamic models, sanding may be performed to ensure that the movable parts will easily mate up and will not bind due to friction from rough surfaces. An additional post-processing step that may need to be taken, depending on the use case of the model, is the addition of magnets, threaded inserts, or other hardware that can be used to connect and hold multiple parts together. 

\section{Results}
\label{sec:headings}

With the use of 3D printing within hospitals for the creation of pre-surgical planning models, patient care has been brought to a whole new level. However, in some surgical cases, monolithic static models may not illustrate the full picture of the anatomy within the area of interest. By utilizing multi-part dynamic models, the patient’s anatomy can be segmented and printed in a manner that fully depicts how the anatomy moves and interacts. By creating dynamic models with multiple parts, physicians can utilize them to see not only how each part aligns with other anatomies, but can then remove the anatomical parts to plan their procedure on a single part of the anatomy without the other anatomies being in the way (Figure 3).

\begin{figure}[h]
    \centering
    \includegraphics[width=0.4\linewidth]{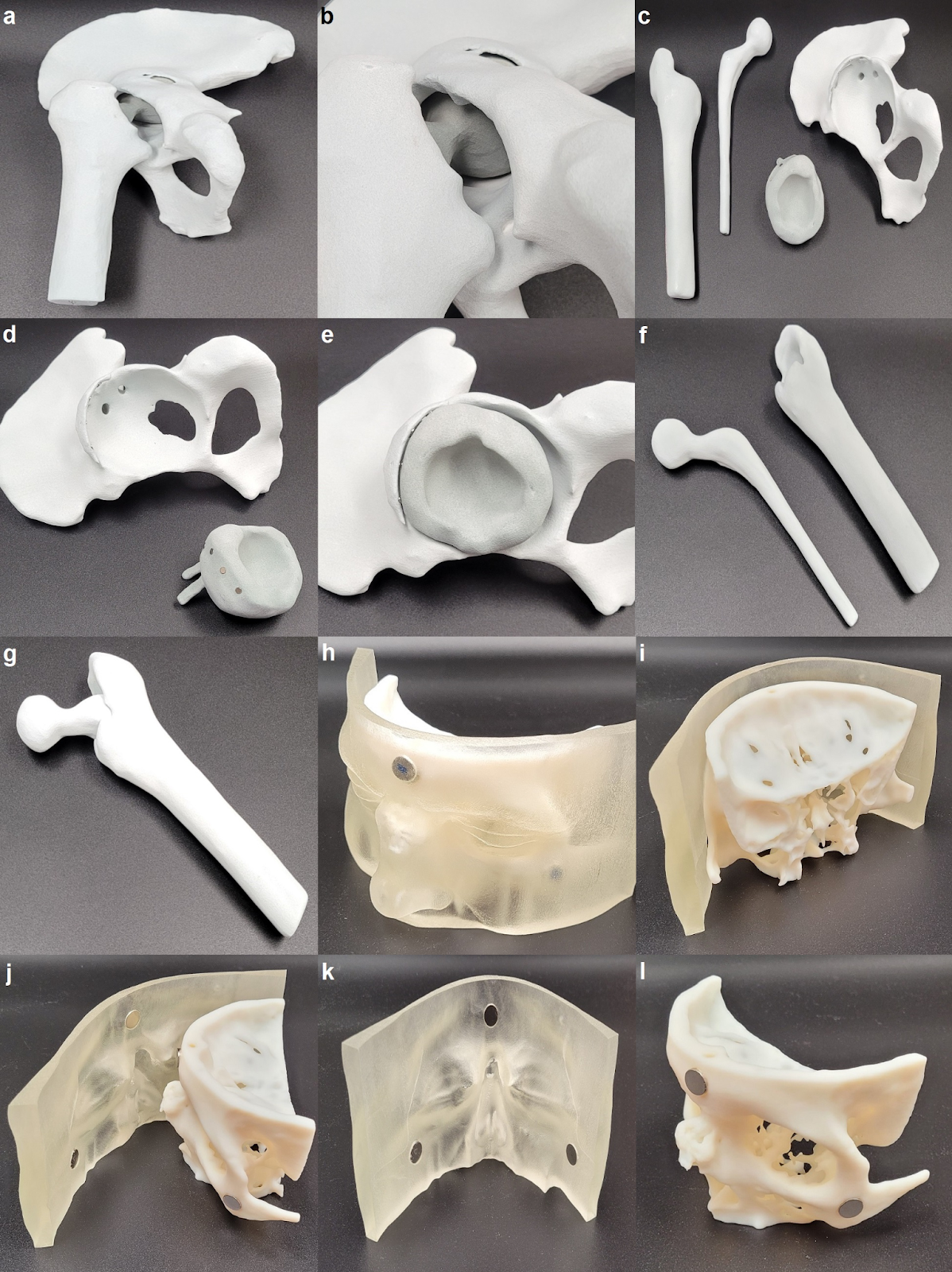}
    \caption{Three unique dynamic models of 3D printed pre-surgical planning models used for hip arthroplasty and nasal bone growth. \textbf{(a)} Dynamic model of a patient’s hemi-pelvis, femur, and femoral prosthesis printed as a single model, which can dynamically move. \textbf{(b)} Close-up of the dynamic joint made up of the hemi-pelvis and femoral prosthesis. \textbf{(c) }Multipart removable pieces for a failed hip arthroplasty model showing the femur, femoral prosthesis, acetabular cup, screws and hemi-pelvis with screw holes. \textbf{(d)} Hemi-pelvis with removable acetabular cup, screws, and added magnets to assist in holding the acetabular cup in the anatomically correct location. \textbf{(e)} Hemi-pelvis model with removable acetabular cup held in place with the modeled screws and magnets. \textbf{(f) }Femur and removable femoral prosthesis models separated into stand alone parts. \textbf{(g) }Femur with removable femoral prosthesis inserted into it depicting how the anatomy is within the patient. \textbf{(h) }Nasal bone growth model including the skull and the patient’s skin which is held in place with magnets. \textbf{(i)} Rear view of the skull and skin attached. \textbf{(j) }Removal of the skin from the skull revealing the nasal bone growth. \textbf{(k)} Rear of the skin which attaches to the skull showing the magnets used to hold the skin model in place. \textbf{(l)} Skull with the skin removed showing the nasal bone growth and the magnets used to hold the skin onto the skull model.
}
    \label{fig:placeholder}
\end{figure}
Although these dynamic models are created in much the same way as a standard pre-surgical planning model, care needs to be taken to ensure that the joints and other dynamic parts within the model do not interfere with each other. By utilizing digital post-processing tools such as smoothing, push \& pull and translate, the parts can be modeled to ensure that there is no interference between moving parts and that removable parts can easily be taken off of the model. One point to take into consideration is that by performing these digital post-processing steps, some parts of the anatomy may not be fully or correctly represented, and the accuracy of the models themselves may differ from the real-world anatomy. Therefore, care should be taken to minimize the number of changes made to the model to ensure the accuracy of the anatomies, or major changes should be communicated to the physician to let them know that there could be differences between the models and the patient’s anatomy.

These dynamic models can be 3D printed using any type of printing technology that is currently available, however, some methods will work better than others and reduce the time needed for post-processing steps. Material jetting and powder bed fusion are two of the best methods to utilize when creating a dynamic model, as the support material is easily removed and the layer lines are thin enough to keep moving parts from interfering and binding with each other. Binder jetting is another beneficial method for the creation of dynamic models; however, the model parts can be quite fragile after removal from the printer and before additional post-processing steps are taken, which could lead to difficulty in making successful dynamic models. Stereolithography and Fused Filament Fabrication can also be used in the manufacture of dynamic models; however, due to the lattice support structures that are used in these methods, along with thicker layer lines, these printing technologies are the weakest choice when it comes to making dynamic models.

As printing technologies continue to evolve, the creation of dynamic models will become easier to produce. Currently, the best printing methods require large and expensive printers that utilize specialized and expensive materials. As these printing methods become cheaper and easier to operate, dynamic models can be produced on a wider scale for more procedures. Another focus for future improvements would be within the software suites that are used for the creation of dynamic models. Additional tools or custom scripts could be developed that would allow the user to quickly and easily ensure that there is no interference between parts, making sure that they can move freely. With relatively minor improvements, on both the hardware and software fronts, dynamic pre-surgical planning models will become easier to create for users of any skill level.

\bibliographystyle{unsrtnat}
\bibliography{references}







\end{document}